\documentclass[lettersize,journal]{IEEEtran}
\IEEEoverridecommandlockouts
\usepackage{cite}
\usepackage{amsmath,amssymb,amsfonts}
\usepackage{algorithmic}
\usepackage{graphicx}
\usepackage{textcomp}
\usepackage{xcolor}
\usepackage{array}
\usepackage[caption=false,font=normalsize,
labelfont=sf,textfont=sf]{subfig}
\usepackage{textcomp}
\usepackage{stfloats}
\usepackage{url}
\usepackage{verbatim}
\usepackage{graphicx}
\usepackage{balance}
\def\BibTeX{{\rm B\kern-.05em{\sc i\kern-.025em b}\kern-.08em
    T\kern-.1667em\lower.7ex\hbox{E}\kern-.125emX}}
\begin{document}
\title{Ion-acoustic solitary waves in a partially degenerate plasma}
\author{Rupak Dey, Gadadhar Banerjee and Amar P. Misra \\
Department of Mathematics, Siksha Bhavana, Visva-Bharati University, Santiniketan-731 235,  India
\thanks{R. Dey thanks University Grants Commission (UGC), Govt. of India, for a junior research fellowship (JRF) with Ref. No. 1161/(CSIR-UGC NET DEC. 2018) and F. no. 16-6 (DEC. 2018)/2019 (NET/CSIR). G. Banerjee also acknowledges financial support from UGC, Govt. of India, under the Dr. D. S. Kothari Post Doctoral Fellowship Scheme (PDF) with Ref. No. F.4-2/2006(BSR)/MA/18-19/0096).   A. P. Misra  wishes to thank   Science and Engineering Research Board (SERB), Government of India,  for a research project with sanction order no. CRG/2018/004475. }}


\maketitle

\begin{abstract}
The propagation  of arbitrary amplitude ion-acoustic   (IA) solitary waves (SWs)  is studied in unmagnetized, collisionless, homogeneous  electron-positron-ion (e-p-i) plasmas with finite temperature degeneracy of both electrons and positrons. Starting from a  set of fluid equations for classical ions and   Fermi-Dirac  distribution for degenerate electrons and positrons, a linear dispersion relation for IA waves is derived. It is   seen that the wave dispersion is significantly modified  due to the presence of positron species and the effects of finite temperature degeneracy of electrons and positrons. In the nonlinear regime,  the Sagdeev's pseudopotential approach is employed to study the existence domain and the evolution of nonlinear IA-SWs  in terms of the  parameters that are associated with the finite temperature degeneracy, the background number densities, and  the thermal energies of electrons and positrons.  It is found that in contrast to classical electron-ion plasmas both the subsonic and supersonic IA-SWs can exist in a partially degenerate e-p-i plasma. 
\end{abstract}

\begin{IEEEkeywords}
Ion-acoustic wave, solitary wave, pseudopotential,  partially degenerate plasma,  electron-positron-ion plasma.
\end{IEEEkeywords}
\section{Introduction}\label{sec-intro}
Electron-positron (e-p) plasmas typically behave as a fully ionized gas composed of two fermions, namely  electrons and positrons having equal mass  but opposite charges. Such plasmas are ubiquitous not only in laboratory  \cite{sarri2015} but also in space and astrophysical environments, e.g., in the early universe \cite{gibbons1983very}, active galactic nuclei bulls \cite{miller1988active}, as well as   white dwarfs \cite{shapiro2008black}. The e-p pair plasmas can be created due to collisions between particles which are accelerated by the electrostatic or electromagnetic waves with or without the influence of the gravitational force. In view of their potential applications in these environments various linear and nonlinear wave phenomena have been studied in e-p plasmas [See, e.g.,  \cite{el2012nonlinear,chatterjee2015,banerjee2020large}]. 
\par 
 The presence of mobile ions in an admixture of  electrons and positrons not only modifies the existing high-frequency wave, but also generates a new low-frequency ion-acoustic (IA) wave mode. So,  the characteristics of nonlinear waves in  e-p  plasmas  significantly differ from those in electron-positron-ion (e-p-i) plasmas due to the presence of these massive ions.   Over the last many years, a number of authors have paid their attention  to investigate the nonlinear propagation of solitary waves  in e-p-i plasmas  using   the Sagdeev potential approach   or the reductive perturbation  technique  \cite{ur2015electrostatic,akbari2010propagation,deeba2012gardner,abdelsalam2008ion, akbari2010effects,misra2011large,baluku2011ion,banerjee2015pseudopotential,popel1995ion}. 
 While the latter is commonly used to study weakly nonlinear small amplitude   electrostatic or electromagnetic perturbations  \cite{misra2012ion,hafez2019nonlinear}, the former, on the other hand, is applicable for the nonlinear evolution of  arbitrary amplitude solitary waves     \cite{banerjee2020large,popel1995ion}.  However, an alternative approach of Sagdeev potential has also been developed by McKenzie and co-workers  \cite{mckenzie2003unified,mckenzie2003electron, mckenzie2002fluid}  where the coherent nonlinear structures are examined in their own frames of reference.  
\par 
The number density of degenerate particles in extremely dense matter  such as cosmic environments, compact astrophysical objects  like  white dwarfs  \cite{boshkayev2016equilibrium} and active galactic nuclei  is so high ($\sim 10^{28}-10^{34}$ cm$^{-3}$) that the average inter particle distance  is considerably smaller than the electron (or positron) thermal de-Broglie   wavelength and so they obey the Fermi Dirac statistics \cite{landau2013course}.  The degenerate pressure,   which depends upon the number density of constituent particles but  independent of its thermodynamic temperature, occurs due to the combined effects of the Pauli's exclusion and  Heisenberg's uncertainty principles.
It has been observed that the said compact objects, which  support themselves against the gravitational collapse by cold, degenerate pressure of fermions (electrons/positrons), are of two types: Type-I consists of those objects (like white dwarfs) which are supported by the pressure of degenerate electrons or positrons  and type-II are those  (like  neutron stars) which are supported by the pressure due to the combination of nucleon degeneracy and nuclear interactions \cite{el2012nonlinear}. The degenerate pressure significantly influences the evolution of electrostatic and electromagnetic perturbations in  degenerate matters \cite{mamun2010arbitrary}. The energy distribution of  degenerate particles in e-p-i plasmas follows  the Fermi-Dirac (FD) distribution  which is usually characterized by two independent parameters: the chemical potential and the thermodynamic temperature. Here, if the thermodynamic temperatures of the electrons $( T_e)$  and positrons $( T_p)$ are comparable to   the corresponding Fermi temperature $(T_{Fj},~ j=e,~p)$ then  the usual Maxwell-Boltzmann distribution is modified to the  FD distribution \cite{landau2013course}.
    In particular,   for   $j$-th species particles when  the condition $T_j  \gg T_{Fj}$ is fulfilled the    particles are then said to  be in the  nondegenerate state  and their background distribution can be governed by the Maxwell-Boltzmann ones.  In the opposite limit, i.e.,  $ T_j  \ll T_{Fj}$ the particles are   completely degenerate and their distributions follow that of the Fermi-Dirac.  However, in real situations, the particle's   temperature $T_j$   may be finite and  not all of them   are degenerate.  In this situation, the partial degeneracy of particles like electrons and positrons  come into the picture in which  case either  $ T_j<T_{Fj} $ or $ T_j>T_{Fj},~ (j=e,~p)$ holds, i.e.,  electrons and positrons are   neither non-degenerate nor  completely degenerate. In the nonrelativistic regime of partial degeneracy, the energy $E$ of electrons and positrons can be considered as $(1/2)mv^2$, where $m$ is the mass and $v$ is the velocity. Since the particles are not completely degenerate there is no strict upper limit of the energy level and one can evaluate the number density and pressure over all the energy levels by extending  the velocity to infinity, i.e.,  $0<v<\infty$. 
\par
On the other hand, the ions are typically non-degenerate as their mass is heavier than that of electrons or positrons.  In the classical limit, the energy distributions of electrons and positrons are described by the Maxwell-Boltzmann distributions which typically depend   on  the thermodynamic temperatures, whereas  in the fully degenerate limit,  the energy distributions depend only on the  chemical potential. So, an intermediate regime exists in which both the parameters, namely the thermodynamic temperature and the chemical potential come into the picture. The present work mainly focuses on this regime of interest.    
 \par 
In view of the  important and interesting consequences in the degenerate regimes many authors have studied    IA wave excitation   in degenerate  plasmas  both  theoretically and experimentally  \cite{misra2021landau,  haas2015linear, popel1995ion, schamel2016correct, iqbal2016nonlinear,hossain2022methods,mannan2022theory,kremp1999quantum,sultana2022review}.  To mention a few,  Haas and Mahmood  \cite{haas2015linear, haas2022linear, haas2016nonlinear} investigated the linear and nonlinear propagation of IA waves in various plasma environments with arbitrary degeneracy of electrons. However, their studies were limited to small-amplitude perturbations and plasmas without the positron species.   The dispersion properties of   electrostatic waves in degenerate plasmas were studied by Melrose \textit{et al.} in the linear regime \cite{melrose2010}.  In an another investigation, it has been found that the presence of hot electron  and positron species gives rise to higher phase velocity, amplitude and width of  small-amplitude  positron acoustic waves in a quantum degenerate plasma  \cite{el2016nonlinear}. In a recent review,  Misra and Brodin \cite{misra2022wave} presented the theoretical background of the wave-particle  interactions and  the  physical mechanism of linear and nonlinear Landau damping in the propagation of electrostatic waves in  degenerate  and non-degenerate quantum plasmas.    Furthermore, the existence of three different kinds of waves and their nonlinear evolution   in degenerate spin polarized plasmas  have been studied by Iqbal and Andreev \cite{iqbal2016nonlinear}.   
\par       
In this paper, our aim is to consider the intermediate degenerate plasma regime,  i.e., in the partially degenerate e-p-i plasmas and study the existence domain and the formation of both  large and small amplitude IA solitary waves in unmagnetized, collisionless, homogeneous e-p-i plasmas with finite temperature degeneracy of both electrons and positrons.  The paper is organized in the following order: In Sec. \ref{sec-model}, we derive the expressions for the number densities of degenerate electrons and positrons in terms of polylogarithmic  function and present the basic set of normalized  equations for   e-p-i plasmas. The linear dispersion relation is derived and analyzed in Sec. \ref{sec-lin}.  The nonlinear analysis is performed in Sec. \ref{sec-nonlin} which comprises  the derivation for the Sagdeev pseudopotential as well as the analyses for large and small amplitude IA waves. Finally, Sec. \ref{sec-conclu} is left to summarize  the results.
\section{Basic Equations} \label{sec-model}
We consider the propagation of IA solitary waves in an e-p-i plasma with finite temperature degeneracy of inertialess electrons and positrons, and classical inertial cold ions.  We assume that  the thermodynamic temperatures of electrons and positrons, to be denoted by,   $T_{j}$   (with the suffix $j=e$ for electrons and $j=p$ for positrons) is slightly larger than their Fermi temperatures $T_{Fj}\equiv E_{Fj}/{K_B}$, i.e., $T_j>T_{Fj}$. Here, $K_B$ is the Boltzmann constant and $E_{Fj}$ is  the Fermi energy for $j$-species   particles.
 In the degenerate regime, the electron and positron number densities $ n_j(\mu_j,T_j)$, and the scalar pressure $p_j(\mu_j,T_j)$ can be obtained using the Fermi-Dirac distribution  as \cite{shukla2011colloquium,eliasson2016finite}
\begin{equation}
n_j=\frac{Li_{3/2}\left[-\exp\left(\xi_{\mu_j}\right)\right]}{Li_{3/2} \left[-\exp\left(\xi_{\mu_{j0}}\right)\right]}, \label{normalized number density}
\end{equation}
\begin{equation}
p_j=\frac{Li_{5/2} \left[-\exp\left(\xi_{\mu_j}\right)\right]}{Li_{3/2} \left[-\exp\left(\xi_{\mu_{j0}}\right)\right]}, \label{normalized scalar pressure}
\end{equation}
where   $Li_{\nu}[z]$ denotes the polylogarithm function of $z$ with index $\nu$. Also,  $ \xi_{\mu_ j} \equiv {\mu_j}/{K_BT_j}$ and $ \xi_{\mu_{ j0}}={\mu_{j0}}/{{K_B}{ T_j}} $ are  the degeneracy parameters corresponding to the perturbed and unperturbed chemical potentials $\mu_j$ and $\mu_{j0}$ respectively. Furthermore,    the number density  $n_j$ and the scalar pressure  $p_j$ are, respectively, normalized  by  the  unperturbed values  $n_{j0}= n_j(\mu_{j0},T_j)$ and   $n_{j0}K_BT_j$ for $ j=e,~ p $. The  Pauli exclusion  principle allows at most one fermion to occupy each possible state, which  in the firm condition (where the kinetic energy, $E\sim0$) results  \cite{misra2021landau}
\begin{equation}
 \sum_{j=e,p} \left[ 1+\exp \left( -\frac{\mu_j}{K_B T_j} \right) \right]^{-1} \leq 1.
\end{equation}
The equilibrium chemical potential  $\mu_{j0}$ is related to the equilibrium density  $n_{j0}$, given by,
\begin{equation}
-\frac{n_{j0}}{Li_{3/2}[-\exp({\xi_{\mu _{j0}})}]}\left(\frac{ m}{2\pi K_B T_j}\right)^{3/2}=2\left(\frac{m}{2\pi\hbar}\right)^3 ,\label{equilibrium number density}
\end{equation}
or the ratio  ${T_{F_j}}/{T_j}$ in which case $\xi_{\mu_{j0}}$ is given by
\begin{equation}
-Li_{3/2}\left[-\exp(\xi_{\mu_{j0}})\right]=\frac{4}{3\sqrt{\pi}}\left(\frac{T_{F_j}}{T_j}\right)^{3/2}.\label{equilibrium degenerate parameter}
\end{equation}
In particular, for  finite temperature degeneracy with $T_j>T_{F_j}$,  Eq. (\ref{equilibrium degenerate parameter}) reduces to
\begin{equation}
\xi_{\mu_{j0}}\approx\ln{ \left[\frac{4}{3\sqrt{\pi}}\left(\frac{T_{F_j}}{T_j}\right)^{3/2}\right]};~j=e,~p. \label{eq-muj0}
\end{equation}
We note that in the non-degenerate limit $(T_j\gg T_{Fj})$, the degeneracy parameter $ \xi_{\mu_ j}$  is large but negative, however, in the case of full degeneracy $(T_j\ll T_{Fj})$, the   parameter $ \xi_{\mu_ j} $ is both large and positive  \cite{shukla2011colloquium}.

\par
Thus, the basic set of normalized  equations governing the dynamics of IA waves consist of the ion continuity and momentum balance equations;   the momentum equations for inertialess electrons and positrons, and the Poisson equation.  These are
 \begin{equation}
 \frac{\partial {n_i}}{\partial t }+\nabla \cdot \left(n_i {\bf u_i}\right)=0,\label{continuity equation}
 \end{equation}
  \begin{equation}
\frac{\partial {{\bf u_i}}}{\partial t }+\left({\bf u_i}\cdot\nabla\right){\bf u_i}=-\frac{1}{\beta_e}\nabla{\phi}, \label{momentum equation of ion}
 \end{equation}
  \begin{equation}
 0=\nabla \phi-\frac{\nabla p _e}{n_e}, \label{momentum equation of electron}
 \end{equation}
 \begin{equation}
 0=\nabla \phi+\sigma\frac{\nabla p_p}{n_p}, \label{momentum equation of positron}
 \end{equation}
\begin{equation}
 \nabla^2 \phi=\beta_e({\alpha_e n_e}-{\alpha_p n_p}-n_i).\label{poisson equation}
 \end{equation} 
Here, $n_i$ is the ion number density  normalized by its unperturbed value $n_{i0}$, $\mathbf{u}_i$ is the ion fluid velocity   normalized by  the generalized IA speed, given by,  \cite{haas2016nonlinear} 
\begin{equation}
c_{s}=   \sqrt{\frac{1}{m_i}\left(\frac{dp_e}{dn_e}\right)_0}= \sqrt{\frac{\beta_e K_BT_e}{m_i}}, \label{ion acoustic speed}
\end{equation}
with $m_i$ denoting the ion mass and $\beta_j=Li_{3/2} \left[-\exp\left(\xi_{\mu_{j0}}\right)\right]/{Li_{1/2} \left[-\exp\left(\xi_{\mu_{j0}}\right)\right]}$, $j=e,~p$. Also, 
 $\phi$ is the electrostatic potential  normalized by   $K_BT_e/e$.  The time and space variables are normalized by the ion plasma period $\omega_{pi}^{-1}=\left({{4\pi n_{i0}e^2}/{m_i}}\right)^{-{1/2}}$  and the Debye length $\lambda_{D}~ (= {c_{s}}/{\omega_{pi}})$  respectively.  Furthermore,  $\sigma ={T_p}/{T_e}$, $\alpha_e={1}/\left({1-\delta}\right)$, $\alpha_p={\delta}/\left({1-\delta}\right) $ with $ \delta={n_{p0}}/{n_{e0}}$. The temperature ratios $\tau_e\equiv T_{Fe}/T_e$   and $\tau_p=T_{Fp}/T_p$ are related by $\tau_p= \delta^{2/3 }\tau_e/\sigma$.  In the non-degenerate  limit, $\exp(\xi_{\mu_{j0}}) \ll 1$ for which $Li_\nu(-\exp(\xi_{\mu_{j0}})) \approx -\exp(\xi_{\mu_{j0}})$ and $\beta_j\approx1$, we obtain  $c_{s}\approx\left({ K_BT_e}/{m_i}\right)^{1/2}$,   $\lambda_{D}\approx \left({k_B T_e}/ {m_i\omega_{pi}^2} \right)^{1/2}$, and from Eq.   \eqref{normalized scalar pressure} the isothermal pressure law: $p_j=n_jK_BT_j$, i.e.,  the well-known classical results are recovered.   
In the opposite case (full  degeneracy)   with $\exp(\xi_{\mu_{j0}}) \gg 1$ for which $Li_\nu(-\exp(\xi_{\mu_{j0}})) \approx -(\xi_{\mu_{j0}})^{\nu}/\Gamma(\nu+1)$, $\mu_{j0}\approx K_BT_{Fj}$ and $\beta_j=(2/3)\tau_j$, one obtains $c_s=\sqrt{(2/3)K_BT_{Fe}/m_i}$,  $\lambda_D=\sqrt{(2/3)K_BT_{Fe}/m_i \omega_{pi}^2}$ and the Fermi pressure law:  $p_j=(2/5)n_{j0}E_{Fj}(n_j/n_{j0})^{5/3}$.  We are, however, interested in the intermediate regime in which $T_j>T_{Fj}$ and $\mu_{j0}$ is determined by Eq. \eqref{eq-muj0}.   
\par  
Next, using Eqs. (\ref{normalized number density}), (\ref{normalized scalar pressure}), (\ref{momentum equation of electron}), and (\ref{momentum equation of positron}), we obtain the following  reduced expressions for the electron  and positron number densities.
\begin{equation} \label{density_electron}
n_e=\frac{Li_{3/2} \left[-\exp\left(\phi+\xi_{\mu_{e0}}\right)\right]}{Li_{3/2} \left[-\exp\left(\xi_{\mu_{e0}}\right)\right]},
\end{equation}
\begin{equation} \label{density_positron}
n_p=\frac{Li_{3/2} \left[-\exp\left(-\frac{1}{\sigma}\phi+\xi_{\mu_{p0}}\right)\right]}{Li_{3/2} \left[-\exp\left(\xi_{\mu_{p0}}\right)\right]}.
\end{equation}
\section{LINEAR THEORY} \label{sec-lin}
\begin{figure} 
\centering
\includegraphics[width=3.4in,height=2.4in]{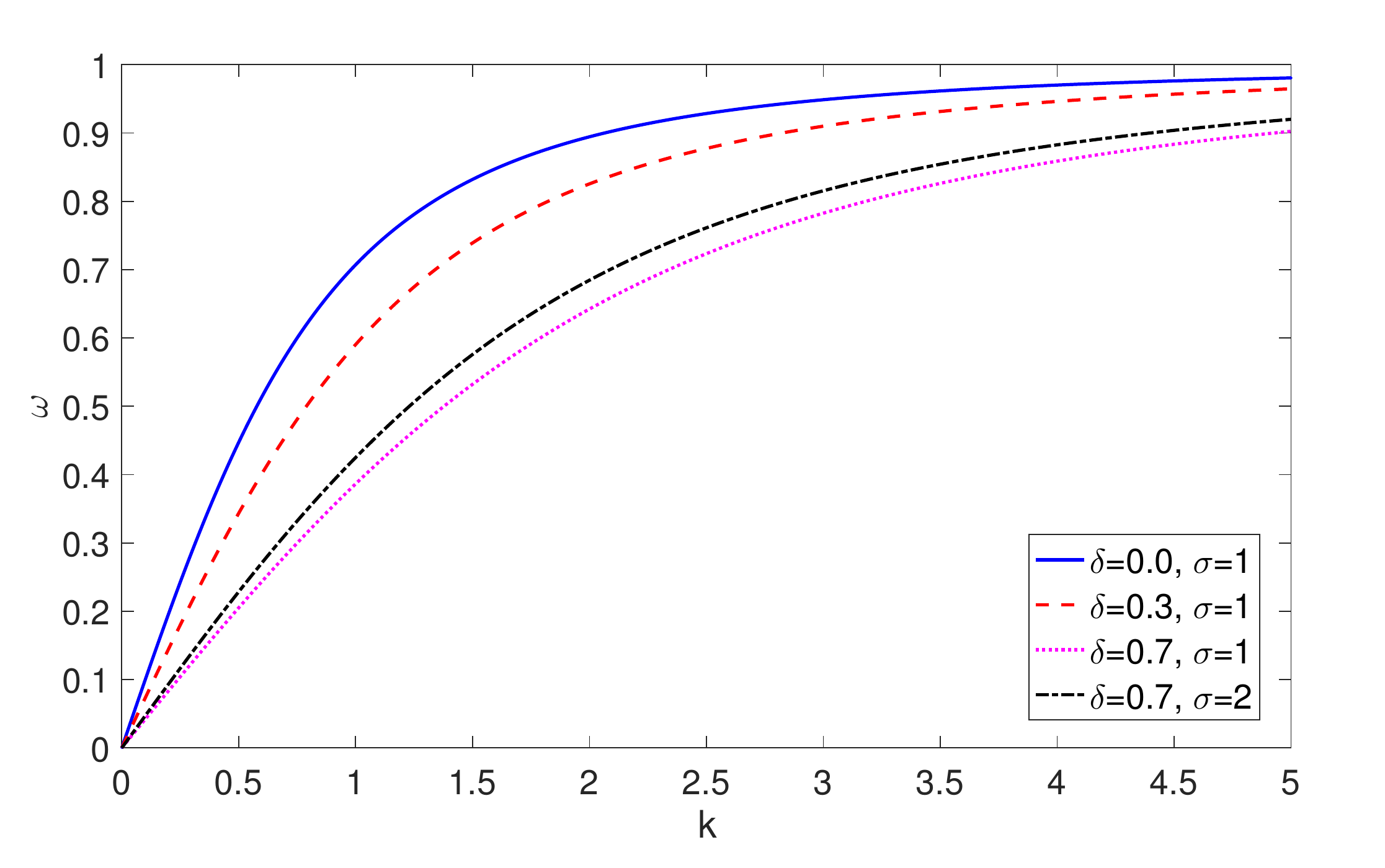}
\caption{ The dispersion relations are plotted for  $\delta=0$  (corresponding to Eq. (\ref{e-i dispersion relation})), $\delta=0.3, 0.7$ (corresponding to Eq. (\ref{e-p-i dispersion relation})). Here, $\sigma=1$ and $T=0.3$.}
\label{fig:disp}
\end{figure}

In order to derive the linear dispersion relation for IA waves, we linearize the system of equations (\ref{continuity equation})-(\ref{poisson equation}) by considering  the first order perturbation quantities relative to the equilibrium values, i.e., 
$n_i=1+n_{i1}$, $n_e=1+n_{e1}$, $n_p=1+n_{p1}$, $\bf {u}_i= \bf{0}+\bf{ u}_{i1} $, $\phi=0+\phi_1$, $\xi_{\mu_e}=\xi_{\mu_{e0}}+\xi_{\mu_{e1}}$, $\xi_{\mu_p}=\xi_{\mu_{p0}}+\xi_{\mu_{p1}}$ etc. Also, we use  the expansion of the polylogarithm function upto the first order, i.e.,
\begin{equation}
\begin{split}
Li_\nu \left[-\exp[{\gamma(x_0+x_1)]}\right]&=Li_\nu\left[-\exp(\gamma x_0)\right]\\
 &+\gamma x_1 Li_{\nu-1}\left[-\exp(\gamma x_0)\right].
 \end{split}
\end{equation}
Assuming the perturbations to vary as plane waves of the form $\sim\exp i({\bf k}\cdot{\bf r} -\omega t)$ with the angular frequency $\omega$ (normalized by $\omega_\mathrm{pi}$)  and the wave vector ${\bf k} $ (normalized by $\lambda_{D}^{-1}$)      we obtain the following   dispersion relation.
\begin{equation}
 \omega^2 =\frac{k^2}{ \alpha_e  +\left({\alpha_p\beta_e}/{\sigma\beta_p}\right) +k^2}. \label{e-p-i dispersion relation}
\end{equation}
Note that the degeneracy  parameter $\beta_e$ is hidden in $k$ due to the normalization with $\lambda_D$. Also,  the ratio $\beta_e/\beta_p$ is more or less than the unity. So, on the linear IA mode,  the effects of the electron and positron degeneracy appear to be insignificant for which the qualitative features of IA waves in partially degenerate e-p-i plasmas will remain the same as in classical e-p-i plasmas \cite{baluku2011ion}. That is, due to the presence of positron species in e-i plasmas as the density ratio $\delta$ increases a significant reduction of the wave frequency and hence that of the phase velocity is seen.  Moreover, in the absence of positron species the dispersion relation (\ref{e-p-i dispersion relation}) exactly agrees with that  reported  in Ref. \cite{haas2022linear}.  The IA waves with such lower   phase velocities may be   weakly
damped or undamped in the wave-particle interactions and thus they can propagate for a longer time. However, the wave frequency can be   increased  with  an increasing value of  the temperature ratio $\sigma$ (See Fig. \ref{fig:disp}).  Physically, increasing the values of  $\delta$ ($\sigma$)   results into an decrease (increase) of the restoring force provided by electrons and positrons which, in turn,   enhances (reduces) the oscillation period or decreases (increases) the wave frequency.
\par  
In particular, by disregarding the contribution of the positron species,  the dispersion equation (\ref{e-p-i dispersion relation}) reduces to the well-known classical form:
\begin{equation}
\omega^2=\frac{k^2}{ 1+k^2}.
\label{e-i dispersion relation}
\end{equation}
\section{NONLINEAR THEORY: SAGDEEV POTENTIAL APPROACH} \label{sec-nonlin}
Going beyond the linear theory, it is pertinent to investigate the nonlinear properties of arbitrary amplitude IA solitary waves in  e-p-i  plasmas with finite temperature degeneracy. To this end, we follow the Sagdeev's pseudopotential approach in which we  assume that all the physical quantities  depend on a single variable $\zeta=l_xx+l_yy+l_zz-Mt$,  
  where $l_x$, $l_y$, and $l_z$ are the direction  cosines of a line along the axes and $M$ is the Mach number (the velocity of the localized wave normalized by $c_s$). Integrating Eqs. (\ref{continuity equation}) and (\ref{momentum equation of ion}), and making use of the boundary conditions, i.e.,     $u_i $, $\phi \to 0$ and $n_i \to 1$ as $\zeta \to \pm\infty$, we obtain the following expression for the ion number density.

\begin{equation}
n_i=\frac{M}{\sqrt{M^2-2\phi/\beta_e}}. \label{ion density}
\end{equation}
Next, substituting the   expressions for the numer densities of electrons, positrons and ions from Eqs. \eqref{density_electron}, (\ref{density_positron}) and (\ref{ion density}) into    Eq. (\ref{poisson equation}), we obtain
\begin{equation}
    \begin{aligned}
   &\frac{d^2\phi}{d\zeta^2}= \beta_e \left[\alpha_e \frac{Li_{3/2} \left[-\exp\left(\phi+\xi_{\mu_{e0}}\right)\right]}{Li_{3/2} \left[-\exp\left(\xi_{\mu_{e0}}\right)\right]}\right. \\
   & \left.-\alpha_p \frac{Li_{3/2} \left[-\exp\left(-\frac{1}{\sigma}\phi+\xi_{\mu_{p0}}\right)\right]}{Li_{3/2} \left[-\exp\left(\xi_{\mu_{p0}}\right)\right]}-\frac{M}{\sqrt{M^2-{2\phi}/{\beta}}} \right]
    \end{aligned}  \label{energy}
    \end{equation}
Finally, integrating Eq. (\ref{energy}) and using the boundary conditions, namely $\phi \to 0$, $ {d\phi}/{d\zeta} \to 0$, and $ {d^2\phi}/{d\zeta^2}\to 0$ as $\zeta \to \pm\infty$, we obtain the following  energy-like equation  for a pseudo particle of unit mass with velocity ${d\phi}/{d\zeta}$ and position $\phi$.   
    \begin{equation}
    \frac{1}{2}\left(\frac{d\phi}{d\zeta}\right)^2+V(\phi)=0,  \label{energy integral}
    \end{equation}
 where the pseudopotential $V(\phi)$ is given by
    \begin{equation}
    \begin{aligned}
    V(\phi)= & \frac{\beta\alpha_e}{Li_{3/2}\left[-\exp(\xi_{\mu_{e0}})   \right]}\\
    &\times \left[Li_{5/2}\left[-\exp{(\xi_{\mu_{e0}})}\right]- Li_{5/2}\left[-\exp{(\phi +\xi_{\mu_{e0}})}\right] \right] \\ 
&    +\frac{\beta\sigma\alpha_p}{Li_{3/2}\left[-\exp(\xi_{\mu_{p0}})   \right]} 
  \left[Li_{5/2}\left[-\exp(\xi_{\mu_{p0}})\right] \right.\\
  &\left. - Li_{5/2}\left[-\exp\left(-\phi/\sigma +\xi_{\mu_{p0}}\right)\right] \right] \\
 &+ \beta^2 M^2 \left[1- \sqrt{1-\frac{2\phi}{\beta M^2}} \right].    
    \end{aligned} \label{Sagdeev Potential}
    \end{equation}
    In what follows, the conditions for the existence of arbitrary amplitude IA solitary waves are \cite{banerjee2015pseudopotential}
    
\begin{itemize}
         \item   $V(0)=V^\prime(0)=0$. 
         \item $V^{\prime\prime}(\phi)< 0$ at $\phi=0$, so that the fixed point at the origin becomes unstable.
         \item $V(\phi_m \neq 0)=0$ and $V^{\prime}(\phi_m)\lessgtr0$ according to when the solitary waves are compressive (with  $\phi>0$) or rarefactive (with   $\phi<0$). Here, $\phi_m $ represents the amplitude of the solitary waves or double layers, if they exist. 
        \end{itemize}
         \par 
         In the following two subsections    \ref{sec-arb-amp} and \ref{sec-small-amp},  we will verify these conditions and study, in details, the domain of the existence of arbitrary amplitude IA solitary waves in the parameter space as well as the  properties of IA solitons  including those in the limit of small amplitude approximation. 
\subsection{ARBITRARY AMPLITUDE WAVE} \label{sec-arb-amp}
 It is easy to verify that the condition (i) is satisfied and the condition (ii) gives  the critical Mach number  $M_l~(<M)$ (i.e., the lower limit of $M$) for the existence of large amplitude solitary waves,   given by, 
\begin{equation}
     M_l=\sqrt{\frac{\sigma\beta_p(1-\delta)}{\sigma\beta_p+ \delta\beta_e}}.
    \label{eq-Ml}
   \end{equation} 
Such critical Mach number must be the same as the phase velocity of linear IA waves to be obtained from Eq. \eqref{e-p-i dispersion relation} in  the limit of $k\ll 1$.  In the non-degenerate limit, Eq. \eqref{eq-Ml} reduces to 
\begin{equation}
M_l \approx \sqrt{\frac{\sigma(1-\delta)}{\sigma+ \delta}}, \label{lower mach}
\end{equation}
 which is in agreement with the results of Ref. \cite{baluku2011ion,popel1995ion}. 
From Eq. \eqref{eq-Ml}, it is clear that the lower limit of the Mach number exists for  $0<\delta<1$ and as   $\delta$   increases in this interval, the value of $M_l$   decreases. In the limit of $\delta \to 0$, we have  $M_l \to 1$, i.e., the nonlinear wave speed equals the ion sound speed. However, the nonlinear IA wave does not exist in the limit of $\delta \to 1$. 
\par 
 Since the pseudopotential  $V(\phi)$ is a real valued function and the polylogarithm function  $Li_{\nu}[z]$  converges for $|z| \leq 1$, we must have the interval for $\phi$: $\phi_c^{-} \leq \phi \leq \phi_c^{+}$, where $\phi_c^{+} =\min \left\lbrace {\beta_e M^2}/{2},\xi_{\mu_{e0}} \right\rbrace$ and $\phi_c^{-}=\sigma \xi_{\mu_{p0}}$. It follows that for a given set of values of $M$ and $\tau_e$ or $\sigma$,  the IA wave amplitude  $\phi_m$ either lies in $0<\phi_m \leq \phi_c^{+}$  or $\phi_c^{-} \leq \phi_m<0$ according to when  the wave potential is positive $(\phi>0)$ or negative $(\phi<0)$. In the interval $\phi_c^{-} \leq \phi_m<0$, there does not exist any local minimum of $V(\phi)$, however, for $0<\phi_m \leq \phi_c^{+}$, it is possible to find a local minimum of $V(\phi)$ for some particular values of the parameters $\delta,~\sigma$, and $\tau_e$ or $\tau_p$ [Since the chemical potentials can be obtained from Eq. \eqref{eq-muj0} in terms of $\tau_j$]. As a consequence,      IA solitary waves or double layers with negative potential do  not exist.    So, we focus on the existence  of solitary waves and/or double layers with positive potential  only. To this end, we need  $V(\phi_c^{+})>0$ and so the upper limit of Mach number $ M_u $ can be obtained by solving  $V(\phi_c^{+})=0$. 
 \par
\begin{figure*} 
\centering
\includegraphics[width=4.5in,height=3in]{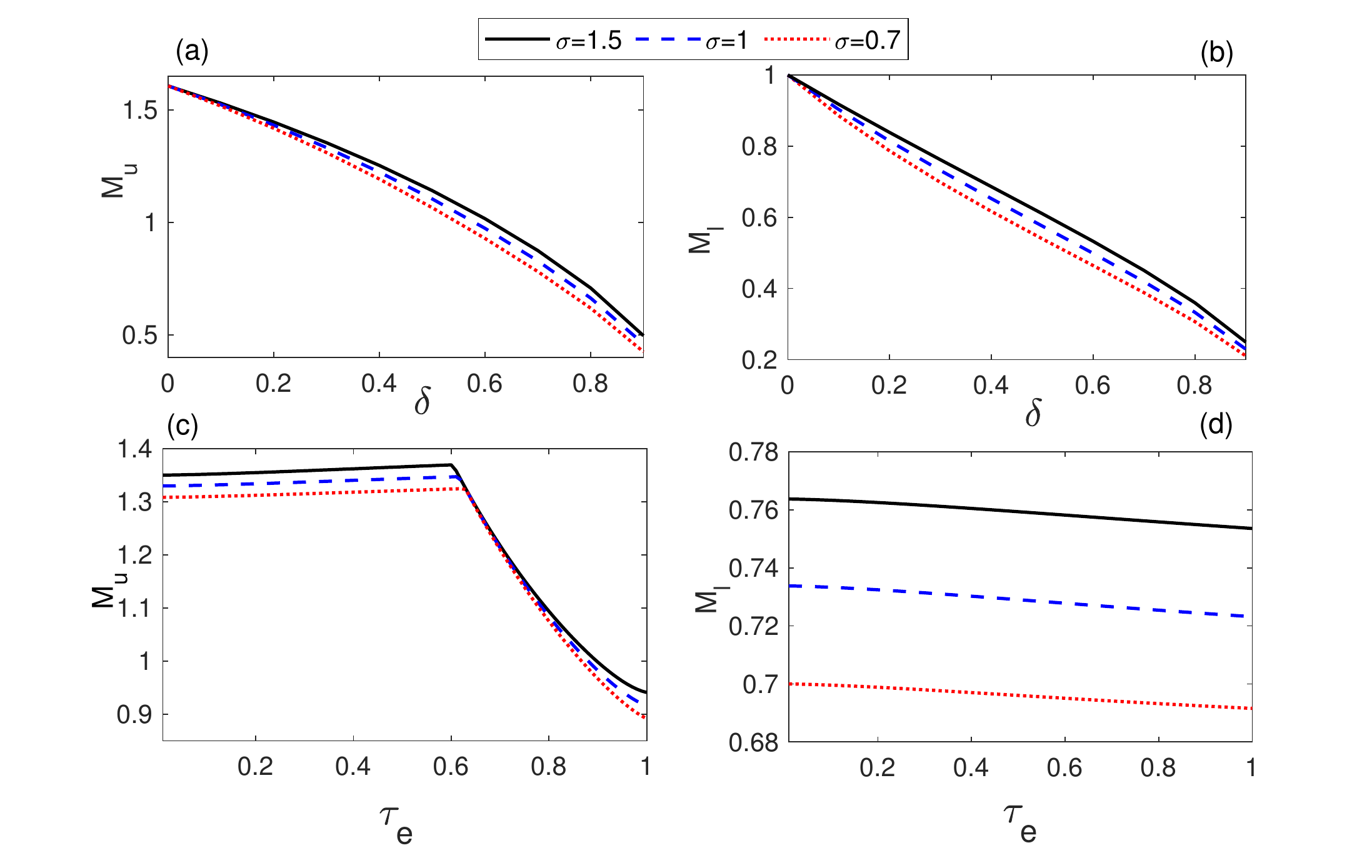}
\caption{ The lower $(M_l)$ and upper $(M_u)$ limits of the Mach number $M$ are plotted against the parameters $\delta$ (with a fixed $\tau_e=0.3$; upper panel) and $\tau_e$ (with a fixed $\delta=0.3$; lower panel)  for three different values of $\sigma$ as in the legend.   }
\label{mach}
\end{figure*}

 
In what follows, we note that for the existence of IA solitary waves we must have the values of $M$ lying in the interval $M_l<M<M_u$. Since an explicit expression of $M_u$ is difficult to obtain,  one can find its values   numerically in terms of the parameters $\delta$, $\tau_e$ or $\tau_p$ and $\sigma$.  The typical values of the  basic plasma parameters can be considered as \cite{manfredi2015} $n_{e0} \sim 2\times 10^{24}$ cm$^{-3}$ and $T_e \sim 10^6 $ K, and use the relations $n_{p0}=\delta n_{e0}$, $T_p=\sigma T_e$, $\tau_e=T_{Fe}/T_e$, and $\tau_{p}=\delta^{2/3}\tau_e/\sigma$ such that $0<\delta<1$ and $0<\tau_e,~\tau_p<1$.  The values of $M_l$ and $M_u$, so obtained, are displayed in Fig. \ref{mach}.  In the latter, the profiles of the Mach numbers are shown against $\delta$ and $\tau_e$ for three different values of $\sigma$. The solid, dashed and dotted curves, respectively,  correspond to the values of $\sigma>1$, $\sigma=1$ and $\sigma<1$. While the values of $M_l$, the lower limit of the Mach number,  are obtained directly from its analytic expression  [Eq. \eqref{eq-Ml}], those of $M_u$ (the upper limit of the Mach number)  are  obtained by a numerical approach.   Figure \ref{mach}  shows that irrespective of the values of $\tau_e~(<1)$ and $\sigma$, both the lower and upper limits of the Mach number tend to decrease below the unity  with   increasing values of $\delta$. As a result,    the range of values of  $M$ for   the existence of IA solitary waves   becomes narrower.   In particular, for $\delta=0$   one finds the range of $M$ as $1\leq M <1.61$ which is in accordance with the well known classical result  $1\leq M <1.585$ for electron-ion plasmas with   Boltzmann distribution of  electrons and  $\tau_e\ll1$. From Fig.   \ref{mach},  an interesting point is to be noted. While the value of $M_l$ decreases slowly with an increasing value of $\tau_e$ in the interval  $0<\tau_e<1$,   there exists a critical value $\tau_{ec}$ of $\tau_e$ such that   $M_u$   increases  slowly   in the interval $0<\tau_e<\tau_{ec}$, however, it decreases in rest of the interval. Thus,   the domain of $M$ for the existence of IA solitary waves increases with increasing value of $\tau_e$ in   $0<\tau_e<\tau_{ec}$.  Moreover, for $\tau_{ec}<\tau_e<1$, since the upper limit of the Mach number starts decreasing,   the existence domain  for $M$  becomes narrower.  The peculiarities of the existence domains of $M$   for all the three cases, namely $\sigma>1$, $\sigma=1$ and $\sigma<1$ remain  similar. From Fig.\ref{mach}, it is also noticed  that the variation of $M$ with $\sigma$ has only a quantitative effect on its  upper and lower limits, i.e.,  both of them increase with increasing values of $\sigma$. Furthermore, inspecting on the regions of the Mach number  we   find   that the IA solitary waves can exist with the Mach numbers  greater than or less than the unity. It turns out that the IA waves can propagate in the form of compressive type solitary waves with  subsonic or supersonic speed depending on the parameter regions we consider.  Also, there exists a critical value of $\delta$ (say $\delta_c$) above which only subsonic solitary waves exist. Such critical value of $\delta$ can be estimated, in particular  for $\sigma=1$ and $\tau_e=0.3$,  as $\delta_c\simeq 0.58$. It is found that in the case of degenerate e-i plasmas with     $\delta\sim0$   only the supersonic IA solitary waves  exist. However,  in  degenerate e-p-i plasmas with $0< \delta \lesssim 0.58$, both the subsonic and supersonic solitary waves, and     only the subsonic solitary waves exist in the regime $0.58\lesssim \delta <1$. 
 \par 
 It is to be mentioned that for the  existence of IA double layers with positive potential, we require, in addition to the   conditions stated in Sec. \ref{sec-nonlin}, the condition  $V(\phi_m)=V^{\prime}(\phi_m)=0$ with $0<\phi_m<\phi^{+}_c$. However, a numerical investigation over a wide range of values of the parameters indicate that   such $\phi=\phi_m$ does not exist in  $0<\phi<\phi^{+}_c$.    The domains of the Mach number with different values of the plasma parameters for which the IA solitary waves exist in partially degenerate e-p-i plasmas are computed as presented in Table \ref{table1}. 

\begin{table*} 
\centering
\begin{tabular}{c c c c c  c c} 
\hline
 $T$ \hspace{0.3cm} & $\delta$ \hspace{0.3cm} & M $(\sigma=0.7<1)$  \hspace{0.3cm} & M $(\sigma=1)$  \hspace{0.3cm} & M $(\sigma=1.5>1)$    \\
\hline
0.1 \hspace{0.3cm} & 0.1 \hspace{0.3cm} & 0.89--1.51 \hspace{0.3cm} & 0.90--1.51 \hspace{0.3cm} & 0.92--1.52  \\
0.1 \hspace{0.3cm} & 0.3 \hspace{0.3cm} & 0.70--1.31 \hspace{0.3cm} & 0.73--1.33 \hspace{0.3cm} & 0.76--1.35  \\
0.3 \hspace{0.3cm} & 0.5 \hspace{0.3cm} & 0.54--1.07 \hspace{0.3cm} & 0.58--1.11 \hspace{0.3cm} & 0.61--1.15  \\
0.3 \hspace{0.3cm} & 0.6 \hspace{0.3cm} & 0.46--0.94 \hspace{0.3cm} & 0.50--0.98 \hspace{0.3cm} & 0.53--1.02  \\
0.5 \hspace{0.3cm} & 0.7 \hspace{0.3cm} & 0.39--0.79 \hspace{0.3cm} & 0.42--0.84 \hspace{0.3cm} & 0.45--0.88  \\
0.5 \hspace{0.3cm} & 0.8 \hspace{0.3cm} & 0.31--0.63 \hspace{0.3cm} & 0.33--0.67 \hspace{0.3cm} & 0.36--0.72  \\
0.6 \hspace{0.3cm} & 0.9 \hspace{0.3cm} & 0.21--0.44 \hspace{0.3cm} & 0.23--0.47 \hspace{0.3cm} & 0.25--0.51  \\
\hline

\end{tabular}

\caption{The domains of the Mach number $(M)$ for the existence of IA solitary waves are presented against different parameter values. }
 \label{table1} 
\end{table*}
 
\par
\begin{figure} 
\centering
\includegraphics[width=3.2in,height=2.5in]{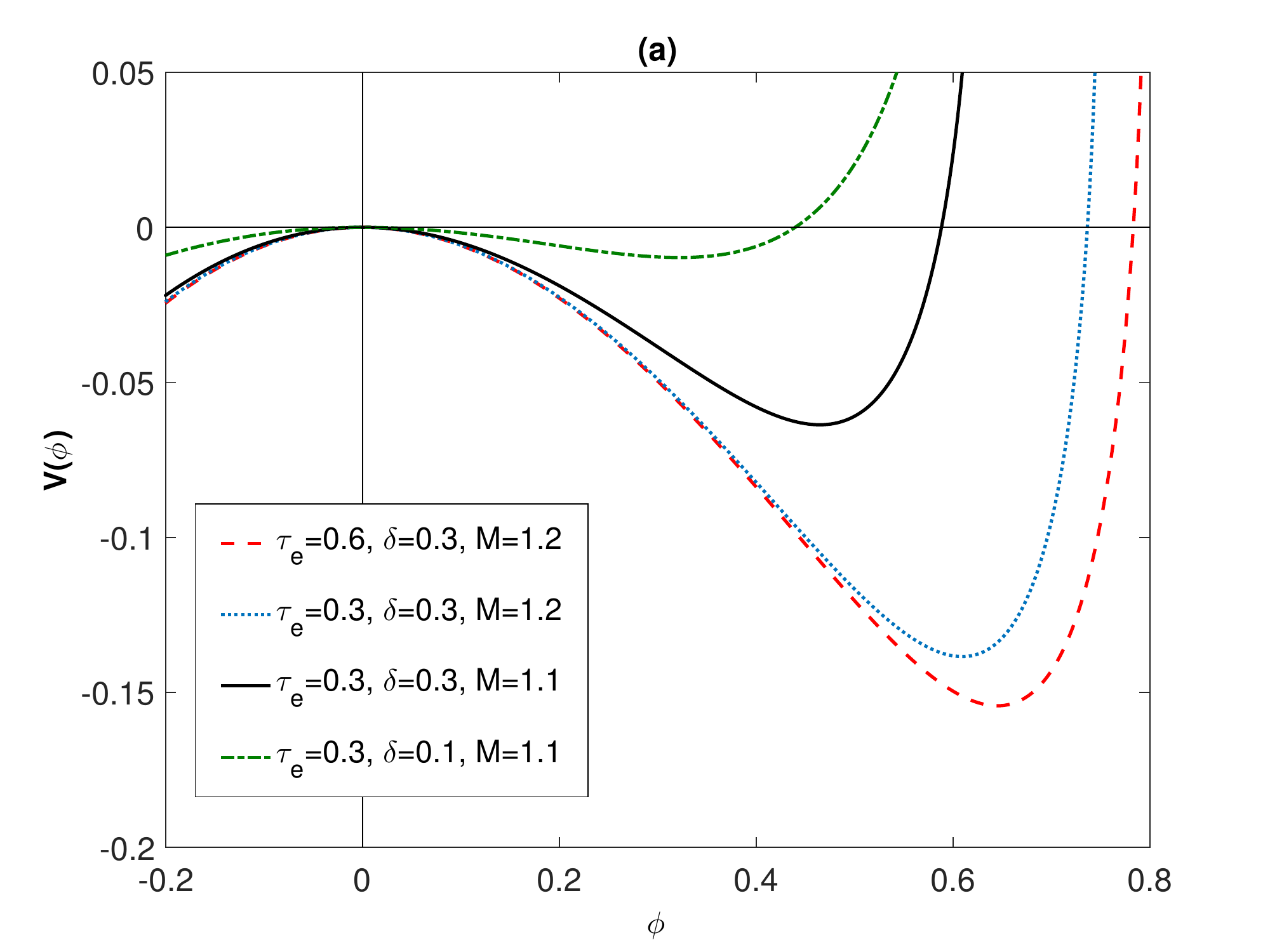}
\includegraphics[width=3.5in,height=2.5in]{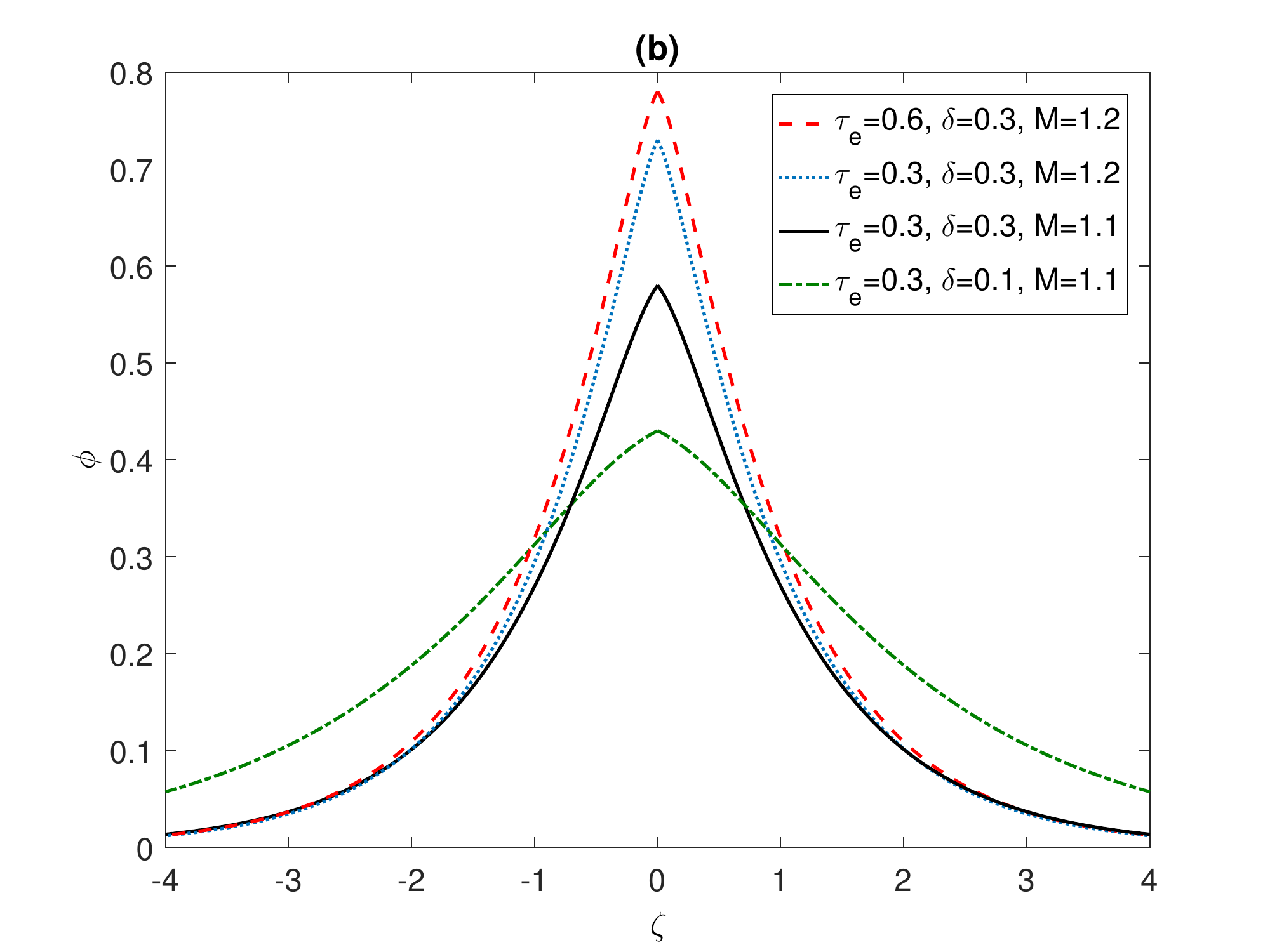}
\caption{ The profiles of the Sagdeev potential $V(\phi)$ [subplot (a)] and the   solitary wave solution $\phi(\zeta)$ are shown with a fixed $\sigma=1$ and  for different values of $\tau_e$, $M$ and $\delta$ as in the legends.  }
\label{fig:pot_sigma_1}
\end{figure}

\begin{figure} 
\centering
\includegraphics[width=3.5in,height=2.5in]{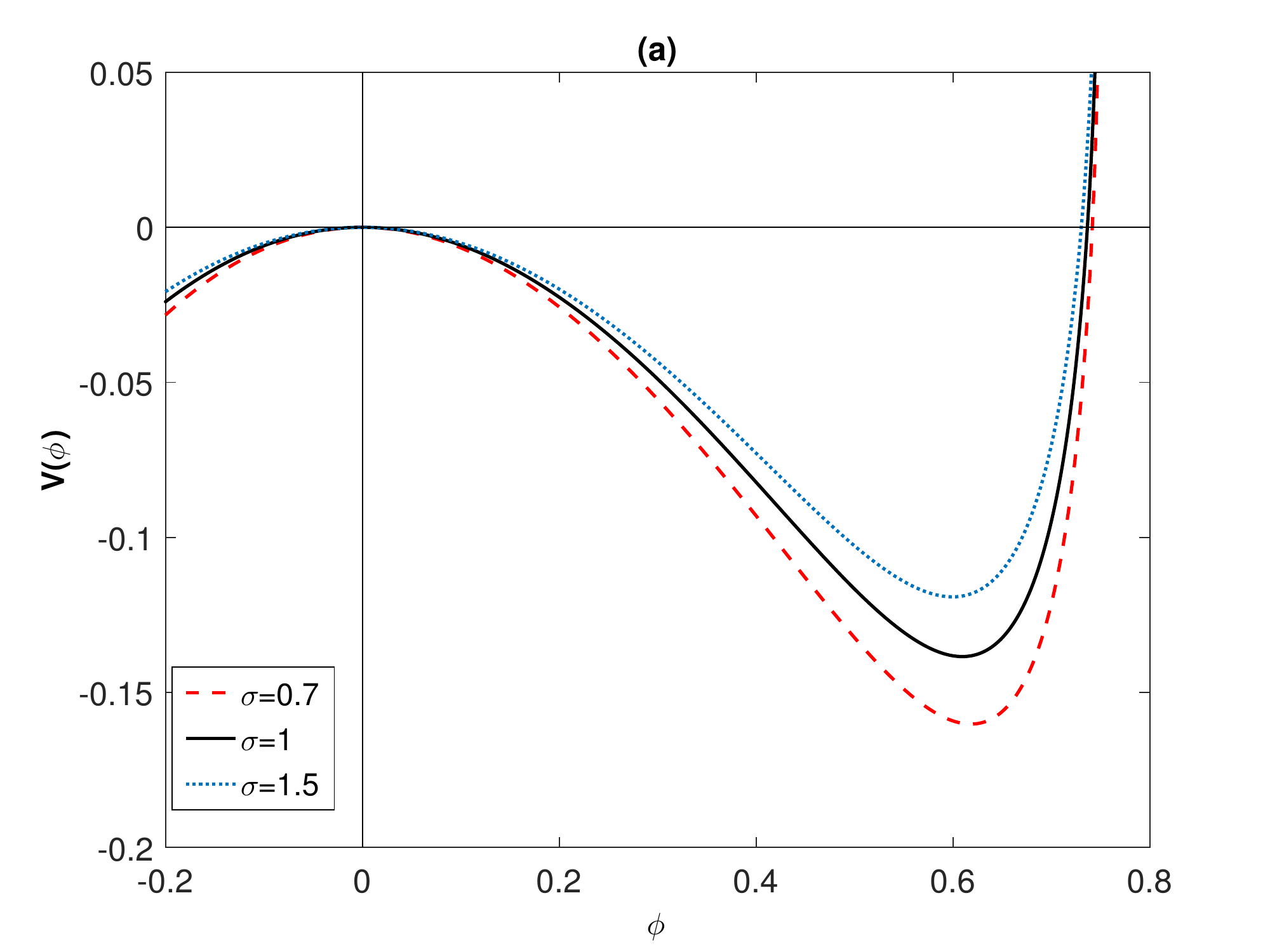}
\includegraphics[width=3.5in,height=2.5in]{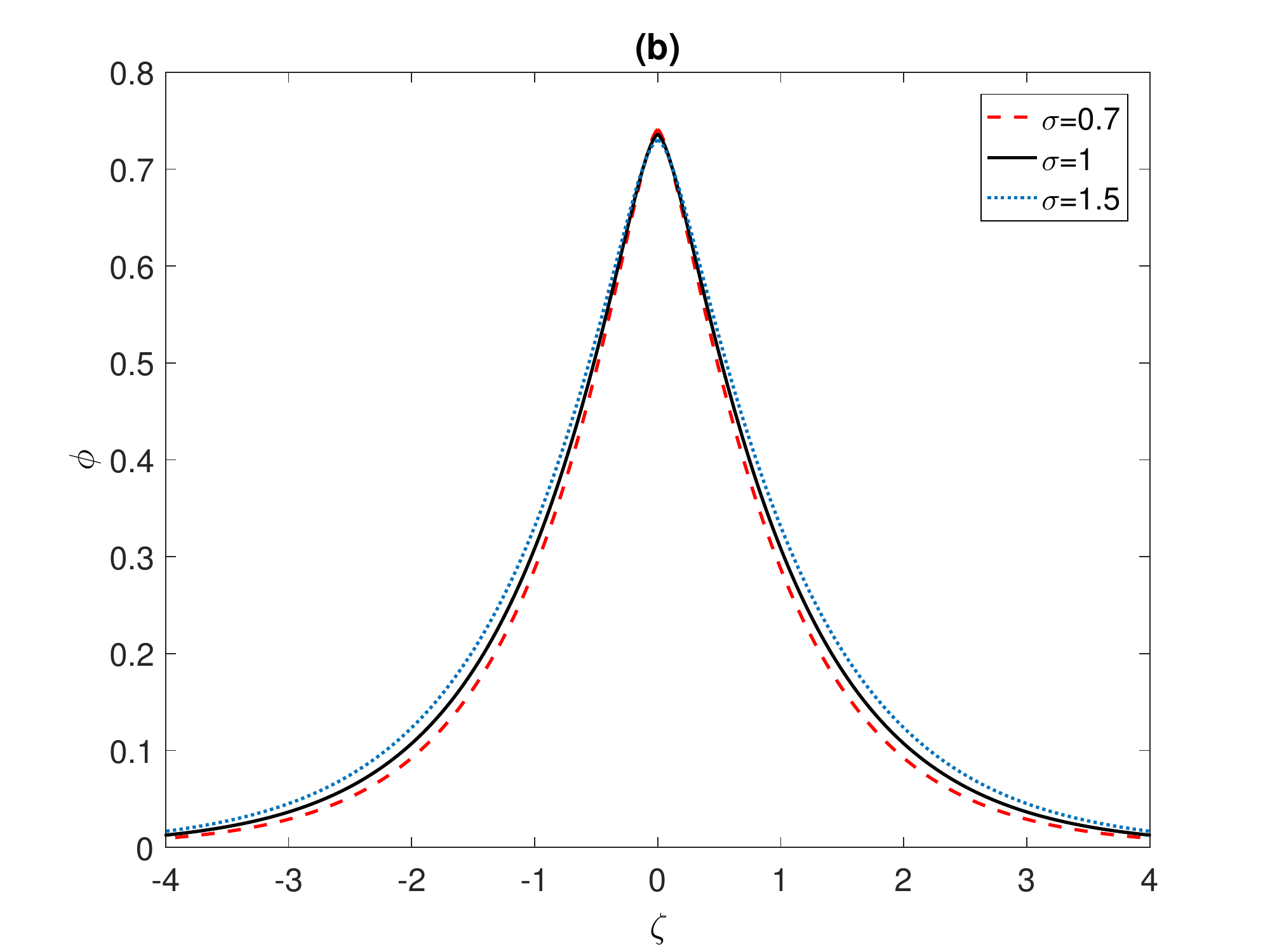}
\caption{ The same as in Fig. \ref{fig:pot_sigma_1} but  for different values of $\sigma$. The other parameter values are fixed at  $M=1.2$, $\delta=0.3$ and $\tau_e=0.7$. }
\label{fig:pot_sigma}
\end{figure}

 
Having obtained various parameter regimes for the existence of IA solitary waves with positive potential, we move on to study the profiles of the solitary waves, especially we focus on the characteristics of the wave amplitudes and widths with the effects of the plasma parameters.  It is, therefore, pedagogic to examine the  profiles of the  Sagdeev potential as well as  the corresponding solitary wave solution of Eq. \eqref{energy integral}. The results are displayed in Figs. \ref{fig:pot_sigma_1} and \ref{fig:pot_sigma}. We find that   for certain parameter values  $V(\phi)$ crosses the $\phi$-axis at $\phi=\phi_m$ and $dV/d\phi<0$ for $0<\phi<\phi^+_c$ (See the upper panels of Figs. \ref{fig:pot_sigma_1} and \ref{fig:pot_sigma}). Such  $\phi_m$ represents the amplitude of the   IA solitary wave which we can verify from the   profiles of $\phi(\zeta)$ (See the lower panels of Figs. \ref{fig:pot_sigma_1} and \ref{fig:pot_sigma}). The width of the solitary wave can be obtained either from the  profile of $\phi(\zeta)$ or by computing the value of $|\phi_m|/|V_\mathrm{min}|$, where  $|V_\mathrm{min}|$ denotes the absolute minimum value of $V(\phi)$.    Here, we  recast the key parameters as $M$, $\tau_e$ or $\tau_p$, $\delta$ and $\sigma$ with   $0<\tau_e,~\tau_p<1$ and   $0<\phi<\phi^+_c$.  For a fixed value of    $\sigma=1$ and different sets of values of $M$, $\tau_e$  and $\delta$ that fall in the existence domain of IA solitary waves,   the  profiles of the pseudopotential   $V(\phi)$ and the corresponding solitary potential    are shown in   Fig. \ref{fig:pot_sigma_1}.  It is found that the depth of the potential well increases having cut-offs at higher values of $\phi$ with an increasing value of each of the parameters $\tau_e$, $M$ and $\delta$ [upper panel of Fig. \ref{fig:pot_sigma_1}].    Such increments are significant even with a small variation of the parameters and are correlated with the increments of both the amplitude and width of the solitary waves [lower panel of Fig. \ref{fig:pot_sigma_1}] except for  the variation with $\delta$ in which case even though the   amplitude of the solitary wave increases, its width gets significantly reduced with a small increment of $\delta$.   
On the other hand,   Fig. \ref{fig:pot_sigma} shows the profiles of $V(\phi)$ (upper panel) and $\phi(\zeta)$ (lower panel) with the variation of the temperature ratio $\sigma$.  It is seen that although the depth of the potential well increases, it has   cut-offs  nearly at the same value of $\phi=\phi_m$ implying that an increase of the wave amplitude is not significant, however, its width increases with increasing values of $\sigma$.   

  \subsection{SMALL AMPLITUDE WAVE } \label{sec-small-amp}
\begin{figure}
\centering
\includegraphics[width=3.5in,height=2.5in]{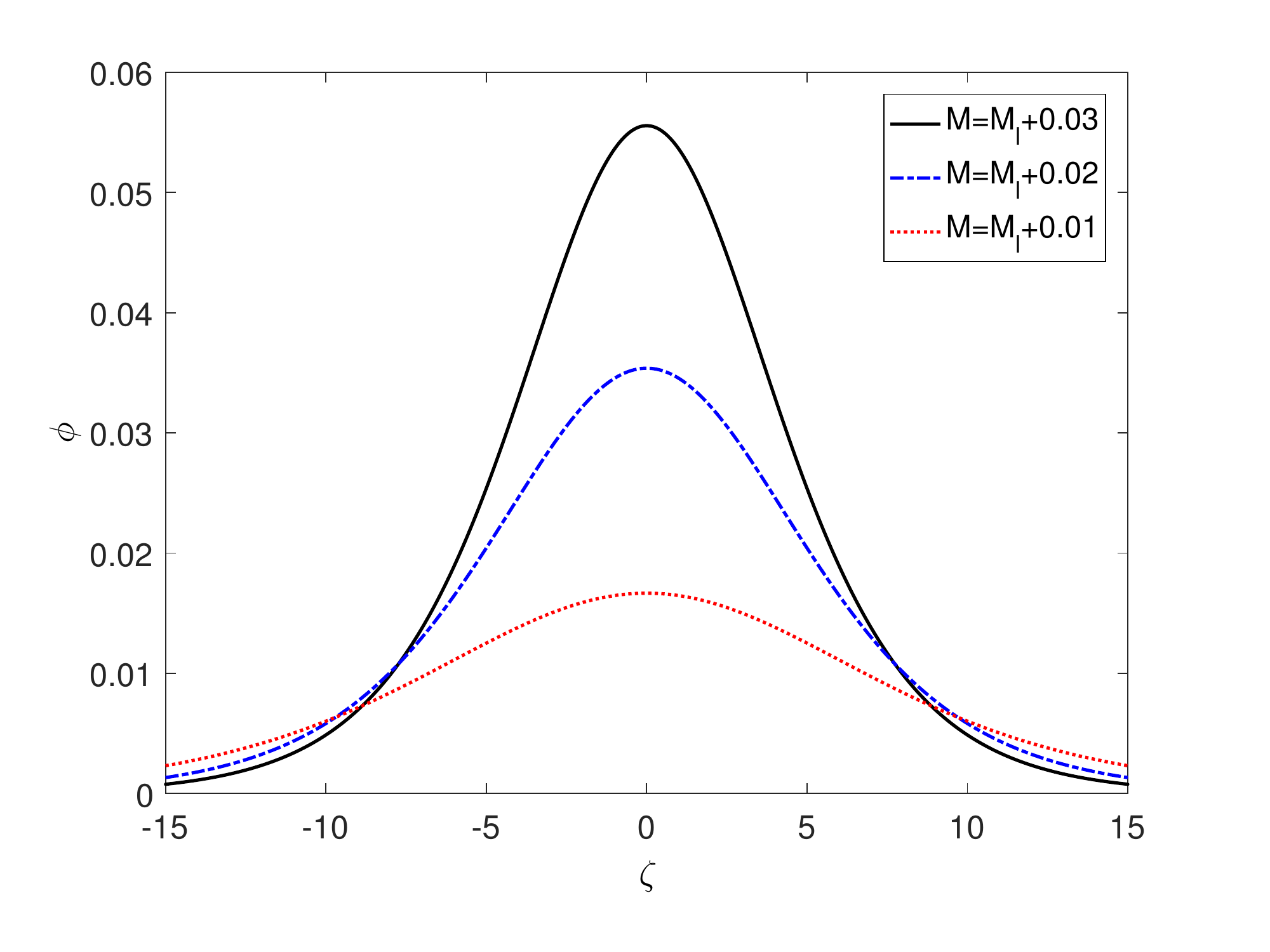}
\caption{ The profiles of the small amplitude IA soliton are shown for different values of the mach number $M$. The other parameter values are $M_l\simeq0.73$, $\tau_e=0.3$, $\delta=0.3$ and $\sigma=1$. The qualitative features of the soliton with variation of other parameters remain the same as shown in Figs. \ref{fig:pot_sigma_1} and \ref{fig:pot_sigma}}.
\label{small_phi}
 \end{figure}

So far we have studied the existence domain and the propagation characteristics of arbitrary amplitude IA solitary waves in terms of different plasma parameters that are relevant in dense plasma environments. One particular interest in this context is to investigate the profiles of  solitary waves when their   amplitudes are no longer arbitrary but small $(|\phi|<1)$. In this case, the pseudopotential $V(\phi)$ can be Taylor expanded about $\phi=0$ up to a reasonable order of $\phi$ so as to obtain a  soliton solution from the energy integral. Thus,  Eq.  \eqref{energy integral} reduces to 
        \begin{equation}
    \frac{1}{2}\left(\frac{d\phi}{d\zeta}\right)^2+ A{\phi^2}+B{\phi^3}=0,  \label{eq-small-V}
    \end{equation}
    where
    \begin{equation}
    \begin{aligned}
   A= \frac{1}{2}&\frac{-\beta_e\alpha_e}{Li_{3/2}\left[-\exp(\xi_{\mu_{e0}})   \right]}{Li_{1/2}\left[-\exp(\xi_{\mu_{e0}})   \right]}\\ 
   &-\frac{\beta_e\alpha_p}{2\sigma Li_{3/2}\left[-\exp(\xi_{\mu_{p0}})   \right]}{Li_{1/2}\left[-\exp(\xi_{\mu_{p0}})   \right]}+\frac{1}{2M^2}
   \end{aligned}
    \end{equation}
     \begin{equation}
    \begin{aligned}
   B=\frac{1}{6} &\frac{-\beta_e\alpha_e}{Li_{3/2}\left[-\exp(\xi_{\mu_{e0}})   \right]}{Li_{-{1/2}}\left[-\exp(\xi_{\mu_{e0}})   \right]}\\ &+\frac{\beta\alpha_p}{6\sigma^2 Li_{3/2}\left[-\exp(\xi_{\mu_{p0}})   \right]}{Li_{-{1/2}}\left[-\exp(\xi_{\mu_{p0}})   \right]}+\frac{1}{2\beta M^4}.
   \end{aligned}
    \end{equation}
Next, integrating Eq. \eqref{eq-small-V} and using the boundary conditions stated before, we obtain the following soliton solution. 
\begin{equation}
  \phi=\phi_m \mathrm{sech}^2\left({\zeta}/{w}\right),
\end{equation} 
where $\phi_m=-A/B$ and $w=(-2/A)^{1/2}$  are, respectively, the amplitude and width of the IA soliton.  Inspecting on the coefficients $A$ and $B$; and the expressions for $\phi_m$ and $w$, we note that for a real soliton solution to exist, one must have  $A<0$. This  leads to the condition $M>M_l$ where $M_l$ is given by Eq. (\ref{minimum mach}). Since the the small amplitude approximation is valid only  for a small deviation from the linear phase velocity, we assume the Mach number as $M=M_l+\varepsilon M_0$, where $\varepsilon$ is some small positive scaling parameter and $M_0$ is the deviation from the linear phase velocity $M_l$.  It is to be noted that at $M=M_l$, the IA waves become  dispersionless and  the corresponding solitary waves   behave like Korteweg-de Vries (KdV)  solitons. Since $A<0$ is noted, the polarity of the soliton depends only on the the sign of $B$.  Thus, the solitary waves with positive (negative) potential exists if  $B>0~(B<0)$. However, the possibility of $B<0$ may be ruled out since in the previous section \ref{sec-arb-amp} we have found only the compressive type solitary waves with arbitrary amplitudes. Furthermore, $B<0$ leads to the condition $M>M_k$, where
\begin{equation}
\begin{split}
      M_k&=\frac{3^{1/4}}{\sqrt{\beta_e}}\left[\alpha_e \frac{Li_{-1/2} \left[-\exp\left(\xi_{\mu_{e0}}\right)\right]}{Li_{3/2} \left[-\exp\left(\xi_{\mu_{e0}}\right)\right]} \right.\\&\left. - \frac{\alpha_p}{\sigma^2}\frac{Li_{-1/2} \left[-\exp\left(\xi_{\mu_{p0}}\right)\right]}{Li_{3/2} \left[-\exp\left(\xi_{\mu_{p0}}\right)\right]} \right]^{-1/4},
   \label{minimum mach}
   \end{split}
   \end{equation}
implying that $M_k-M_l<\varepsilon M_0$. However, there is no parameter region for which  this condition may be satisfied.   So, the propagation of  small amplitude IA solitary waves with negative potential  may not exist. On the other hand, choosing some $M_0$ satisfying $\varepsilon M_0<M_k-M_l$, the propagation of  small amplitude IA solitary waves with positive potential is possible. The qualitative features of the solitons with the variations of the parameters  will remain the same as for arbitrary amplitude waves. However, since the Mach number has some deviation  from $M_l$, its effects (with some new values) on the soliton profiles  are to be noticed (See Fig.  \ref{small_phi}). It is found that in contrast to the large amplitude waves, as the Mach number slightly increases the amplitude of the soliton decreases but its width increases.  

\section{SUMMARY AND CONCLUSION} \label{sec-conclu}
We have studied the propagation characteristics of arbitrary amplitude solitary waves in a multi-component  plasma with finite temperature degeneracy of both electrons and positrons, and classical ions. Starting from a set of fluid equations for classical ions and Fermi-Dirac distribution for electrons and positrons  a linear dispersion relation for IA waves is derived. It is seen that the wave frequency is significantly reduced due to the presence of positron species, however, it can be increased by increasing the relative temperature of positrons compared to electrons. On the other hand, the nonlinear theory of IA solitary waves is studied   using the Sagdeev's pseudopotential approach.   Different  domains  in parameter space for the existence of IA solitary waves are obtained and analyzed. While the lower limit of the Mach number $M_l$ is obtained analytically    from the Sagdeev's condition $(ii)$, its upper limit   $M_u$ is obtained  numerically from the condition $V(\phi_c^{+})>0$, where $\phi_c^{+} =\min \left\lbrace {\beta_e M^2}/{2},~\xi_{\mu_{e0}} \right\rbrace$. Although, the degeneracy effect has no direct influence on the linear wave mode, it expands the existence domain of the Mach number significantly until  $T_{Fj}/T_j\lesssim0.6$; $j=e,~p$. In contrast, the   presence of the positron species reduces the existence regions of $M$ with   increasing the density ratio $\delta$.  The main results can be summarized as follows: 
\begin{itemize}
\item[(a)] The wave frequency and hence the phase velocity of IA waves is reduced due to the presence of positron species. However,  it can be increased with an increase of the positron to electron temperature ratio. Such a reduction of the wave frequency may be desirable for IA waves not to be strongly damped in wave-particle interactions.    
\item[(b)] A numerical investigation indicates that the ion-acoustic double layers do not exist. However,  the solitary waves exist only of the compressive type and in contrast to classical electron-ion plasmas they can propagate  with subsonic or supersonic speed depending on the parameter regimes we consider. Typically, for $\delta\equiv n_{p0}/n_{e0}\sim0.1$, $\tau_e\equiv T_{Fe}/T_e\sim0.1$, the domain of $M$ can be estimated as $0.9\lesssim M\lesssim 1.5$, or, in dimension,  $8\times 10^6~\mathrm{cm/s}\lesssim M\lesssim 1.35\times 10^7~\mathrm{cm/s}$ for $T_e\sim T_p\sim10^6$ K.
\item[(c)] The lower $(M_l)$ and upper $(M_u)$ limits of the Mach number  are highly dependent on the positron to electron number density $\delta$ and the degeneracy parameter $\tau_e$ or $\tau_p$. Their values decrease with increasing values of $\delta$, thereby reducing   the  domain of $M$ for the existence of IA solitary waves.  In the limits of  $\delta \rightarrow 0$ and $\tau_e,~\tau_p\rightarrow 0$, the well-known classical results for electron-ion plasmas are recovered, i.e.,   $1<M<1.585$.  
\item[(d)] Due to finite temperature degeneracy of electrons and positrons, there exists a critical value of $\tau_e$ or $\tau_p$  where the upper limit of the Mach number $M_u$ can turn over, tend to decrease and then close to the value of $M_l$.   
\item[(e)] Both the amplitude and width of the solitary waves are increased due to  enhancements of $M$  and $\tau_e$ or $\tau_p$. However, the amplitude increases but the width decreases with increasing values of $\delta$.  
\end{itemize}
\par 
To conclude,  multi-component degenerate plamas are not only ubiquitous in astrophysical environments, but also   in inertial confinement fusion (ICF) experiments with particle density ranging from $10^{24}$ to $10^{27}~\mathrm{cm}^{-3}$ and the thermodynamic temperature $T_e\sim T_p\sim 10^6-10^8~\mathrm{K}$ \cite{manfredi2015}.  Also, during  ultra-intense short pulse laser irradiation of solid density targets, quantum statistical effects tend to become more prominent than mechanical (quantum diffraction) effects. Furthermore, laboratory simulation of   astrophysical  phenomena in dense plasma environments better fits with the intermediate classical-quantum regime \cite{cross2014}. Thus, our theoretical results should be useful for understanding the localization of arbitrary amplitude ion-acoustic solitary waves that are candidates in these environments. 

\bibliographystyle{IEEEtran}
\bibliography{ref.bib}{} 
 \begin{IEEEbiography}[{\includegraphics[width=1.0in,height=1.0in,clip,keepaspectratio]{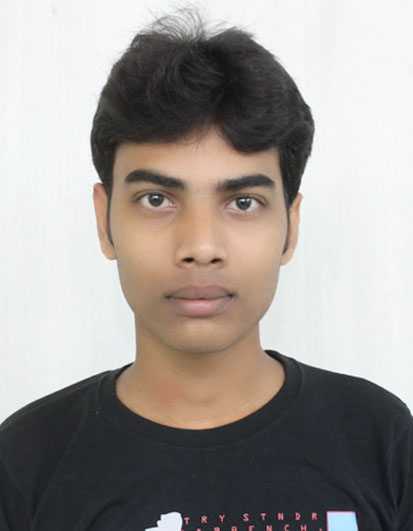}}]{Rupak Dey} is currently a Ph. D. student  in the Department of Mathematics, Siksha Bhavana, Visva-Bharati  University,   India. He received his B. Sc. in  Mathematics from   Vidyasagar University, India in 2015 and M. Sc. in Mathematics from University of Calcutta, India in 2017. His current research interests include nonlinear wave dynamics in plasmas, finite temperature degenerate plasmas and relativistic plasmas. 
\end{IEEEbiography} 
  \begin{IEEEbiography}[{\includegraphics[width=1.0in,height=1.2in,clip,keepaspectratio]{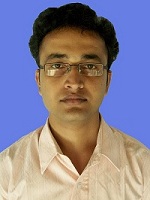}}]{Gadadhar Banerjee} received his B. Sc. in  Mathematics from The University of Burdwan, India in 2008.  He received his M.Sc. and PhD degree from Indian Institute of Technology (IIT) Kharagpur and National Institute of Technology (NIT) Durgapur in 2010 and 2018  respectively. He is currently on leave from Department of Basic Science and Humanities, University of Engineering and Management (UEM), Kolkata, India and working as a post-doctoral research fellow in Department of Mathematics,   Visva-Bharati   University,   India. He is a life member of Plasma Science Society of India (PSSI). He has authored various research papers in reputed journals in the field of theoretical plasmas physics which involves strongly coupled plasmas, dusty plasmas, and quantum plasmas. His current research interests
include nonlinear waves and instabilities   in plasmas, finite temperature degenerate plasmas, relativistic plasmas and nonlinear thermoacoustic waves in plasmas.  
\end{IEEEbiography}  
\begin{IEEEbiography}[{\includegraphics[width=1in,height=1.6in,clip,keepaspectratio]{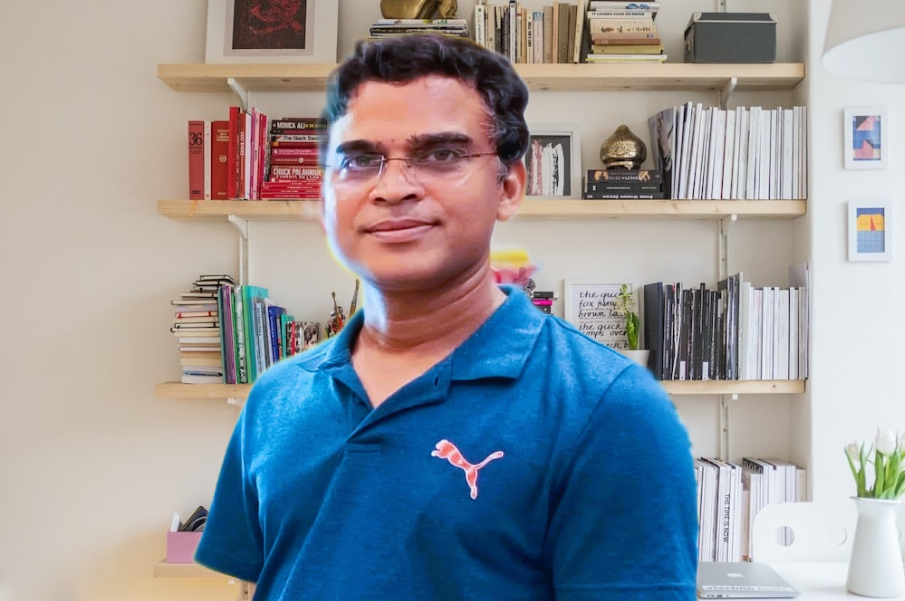}}] {Amar Prasad Misra}
  received the B. Sc. degree in Mathematics in 1995 and M. Sc. degree   in Applied Mathematics in 1997  from  University of Calcutta, India, and the Ph. D. degree in 2006 from (Department of Physics),  Jadavpur University, India. He was a postdoctoral fellow (2009-2011) in the Department of Physics,  Ume{\aa} University, Sweden. He has been a faculty member in the Department of Mathematics, Visva-Bharati University, India since 2005. He is a life member of Plasma Science Society of India (PSSI) and Association of Asia Pacific Physical Societies-Division of Plasma Physics (AAPPS-DPP). He has published more than $100$ research papers in peer reviewed international journals of repute.  He has been a member of the editorial boards  of Physica Scripta since 2019 and Annals of Applied Sciences since 2021. He is an Associate Editor (for Low-Temperature Plasma Physics) of Frontiers in Physics and Frontiers in Astronomy and Space Sciences. His current research interests focus on nonlinear waves and instabilities in plasmas, atmospheric waves, magnetohydrodynamics, chaos in nonlinear dynamical systems, chaos based cryptography and network security.
  \end{IEEEbiography}

\end{document}